# Calculation Binding energy for $^{223,225,227}$Ra isotopes in relativistic heavy cluster model

Keivan Darooyi Divshali† and Mohammad Reza Shojaei

*Faculty of Physics and Nuclear Engineering, Shahrood University of Technology, P.O. Box 36155-316, Shahrood, Iran.*

Abstract:

*$^{14}$C is a stable isotope and emitted from medium and heavy mass nuclei. The $^{14}$C result is in excellent agreement with a favored ground-state-to-ground-state transition according to the cluster model of Blendowske et al. We study Ra isotopes in relativistic core-cluster model, solve the Dirac equation with the new phenomenological potential by parametric Nikiforov-Uvarov method, and obtain wave function and binding energy.*

## 1. Introduction

Since the discovery of $^{14}$C radioactivity of $^{223}$Ra by Rose and Jones [1], the spontaneous emission of neutron-rich clusters, ranging from $^{14}$C to $^{34}$Si, have been observed in the radioactive decay of various trans-lead nuclei [2]. The α cluster structures have been extensively studied over the years for example ref [3-5], Since the binding energy per nucleon is extremely large in $^4$He, it can be a building block of the nuclear systems called α cluster. Therefore, we can give a different isotope as a cluster in the study of medium and heavy mass nuclei. One of the issue we can discuss about that, it is the possibility of the $^{14}$C nucleus could be a cluster, which reviews a few reasons in the following arguments:

$^{14}$C is strongly bound, this is because the proton number 6 corresponds to the sub-closure of the $p_{3/2}$ subshell of the jj-coupling shell model, and the neutron number 8 is the magic number corresponding to the closure of the p shell. Although $^{14}$C has two valence neutrons, the lowest threshold to emit particle is the neutron threshold at $E_x$ = 8.18 MeV, which is high enough value. Also because of strong shell effects, there is no excited state below $E_x$ = 6 MeV just like in $^{16}$O[6]. The β-decay of free $^{14}$C is very slow reflecting the stability of this nucleus. Thus, it is a famous nucleus used for age determination [7] and the half-life of $^{14}$C is 5700 years.

The second argument is that $^{14}$C emitted from some heavy nuclei in the process that is call cluster decay. In reality, there is much more experimental data for the $^{14}$C emission as compared with the emission of $^{12}$C in medium and heavy mass nuclei, and neutron richness of $^{14}$C can be important for that[8]. These experimental data also strongly suggest that $^{14}$C can be a building block of cluster structures in heavy nuclei. $^{14}$C is a favorite clustering in neutron-rich nuclei, even in light nuclei for example in ref [9] the cluster model in O isotopes studied. Figure1 is a schematic of the model assumed.

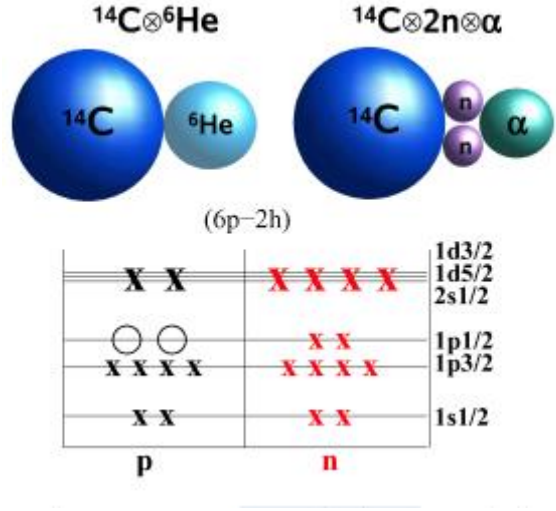

Figure1. Schematic illustration of two possible clusters and shell model structures in $^{20}$O [9].

We study $^{223,225,227}$Ra isotopes in a relativistic core-cluster model with $^{14}$C heavy cluster. For Ra isotopes, we solve the Dirac equation with new phenomenological potential by the parametric Nikiforov-Uvarov (PNU) method. In $^{223}$Ra propose $^{223}$Ra→$^{209}$Pb+$^{14}$C where $^{209}$Pb with 3.25-hour half-life is our core and $^{14}$C is heavy cluster and for $^{225,227}$Ra listed in Table1.

## 2. Wave function and Energy equation

The wave function of the Dirac equation as follows:

$$\Psi_{n_r,k} = \frac{1}{r}\begin{pmatrix} F_{n_r,k}(r)Y_{jm}^l(\theta,\varphi) \\ iG_{n_r,k}(r)Y_{jm}^l(\theta,\varphi) \end{pmatrix} \qquad (1)$$

The radial part of the Dirac equation as follows:

$$\begin{cases} \left(\dfrac{d}{dr}+\dfrac{k}{r}\right)F_{n_r,k}(r) = \dfrac{1}{\hbar c}\left(Mc^2 + E - \Delta(r)\right)G_{n_r,k}(r) \\ \left(\dfrac{d}{dr}-\dfrac{k}{r}\right)G_{n_r,k}(r) = \dfrac{1}{\hbar c}\left(Mc^2 - E + \Delta(r)\right)F_{n_r,k}(r) \end{cases} \qquad (2)$$

For spin symmetry ($\Delta(r)=0$) [10]:

$$\left(-\frac{d^2}{dr^2} + \frac{k(k+1)}{r^2} + \frac{1}{\hbar^2 c^2}\left(Mc^2+E\right)\left(Mc^2-E+\Sigma(r)\right)\right)F_{n_r,k}(r) = 0 \qquad (3)$$

The core-cluster interaction is describing as follows:

$$\Sigma(r) = -v_0 \tanh(\alpha r) - \frac{v_1}{\cosh^2(\alpha r)} - \frac{v_2(1+\cosh(\alpha R))}{\cosh^2(\alpha r)} + \frac{v_3 e^{-\alpha r}}{\cosh(\alpha r)} \qquad (4)$$

Where $v_0, v_1, v_2$ and $v_3$ are the depth of the potential well, R is the nuclear radius and $\alpha$ is related to the range of the potential, You can see the behavior of this potential in Figure 2. We replace (4) in (3), to solve the Dirac equation by the PNU method, it must convert into the following general form.

$$\left(\frac{d^2}{ds^2} + \frac{c_1 - c_2 s}{s(1-c_3 s)}\frac{d}{ds} + \frac{(-p_2 s^2 + p_1 s - p_0)}{s^2(1-c_3 s)^2}\right)\Psi_n(s) = 0 \qquad (5)$$

If we can write the Dirac equation as above, the wave function and the energy equation respectively obtain from the following equations

$$\Psi_{n,k}(s) = N_{n,k} s^{c_{12}} (1-c_3 s)^{c_{13}} P_n^{(c_{10},c_{11})}(1-2c_3 s), \qquad (6)$$

$$nc_2 - (2n+1)c_5 + (2n+1)(\sqrt{c_9} + c_3\sqrt{c_8}) + n(n-1)c_3 + c_7 + 2c_3 c_8 + 2\sqrt{c_8 c_9} = 0, \qquad (7)$$

You can for more details on parameters, see ref [11, 12].

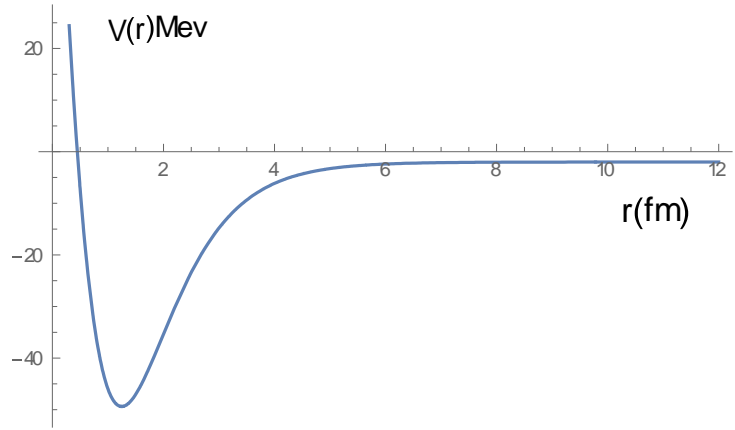

Figure 2. The behavior of potential

Back to spin symmetry Dirac equation, for $1/r^2$ can use the Pekeris-type approximation

$$\frac{1}{r^2} = \frac{1}{R^2}\left(A_0 + \frac{A_1}{1+e^{-2\alpha r}} + \frac{A_2}{(1+e^{-2\alpha r})^2}\right) \qquad (8)$$

Also, select

$$s = (1+e^{-2\alpha r})^{-1} \qquad (9)$$

Eventually, we have

$$\left[\frac{d^2}{ds^2} + \frac{1-2s}{(1-s)s}\frac{d}{ds} + \frac{(2\alpha)^{-2}}{(1-s)^2(s)^2}\left((a_3-a_4)s^2 + (a_1-a_2+a_4)s + a_0+a_2\right)\right]F(s) = 0 \quad (10)$$

Where

$$a_0 = -\frac{(Mc^2+E_{n,k})}{\hbar^2c^2}(Mc^2-E_{n,k}) - \frac{k(k+1)A_0}{R^2} \quad (11)$$

$$a_1 = -\frac{k(k+1)A_1}{R^2} + v_0\left(\frac{Mc^2+E_{n,k}}{\hbar^2c^2}\right), \quad (12)$$

$$a_2 = (v_0+2v_3)\left(\frac{Mc^2+E_{n,k}}{\hbar^2c^2}\right), \quad (13)$$

$$a_3 = -\frac{k(k+1)A_2}{R^2}, \quad (14)$$

$$a_4 = -(4v_1+2v_2(1+\cosh(\alpha R)))\left(\frac{Mc^2+E_{n,k}}{\hbar^2c^2}\right) \quad (15)$$

Finally, we obtain the wave function and energy equation in Eqs. (16) and (17) respectively.

$$F_{n_r,k} = N_{n,k}(1+e^{-2\alpha r})^{-\sqrt{\frac{(Mc^2+E_{n_r,k})}{\hbar^2c^2}(Mc^2-E_{n_r,k}+V_0+2V_3)+\frac{k(k+1)A_0}{R^2}}}$$

$$\times\left(1-\frac{1}{1+e^{-2\alpha r}}\right)^{\sqrt{\frac{k(k+1)}{R^2}(A_2+A_1+A_0)+\frac{(Mc^2+E_{n_r,k})}{\hbar^2c^2}\left((Mc^2-E_{n_r,k}+V_0+4V_3)\right)}} \quad (16)$$

$$\times P_n^{\left(2\sqrt{\frac{(Mc^2+E_{n_r,k})}{\hbar^2c^2}(Mc^2-E_{n_r,k}+V_0+2V_3)+\frac{k(k+1)A_0}{R^2}},\, 2\sqrt{\frac{k(k+1)}{R^2}(A_2+A_1+A_0)+\frac{(Mc^2+E_{n_r,k})}{\hbar^2c^2}\left((Mc^2-E_{n_r,k}+V_0+4V_3)\right)}\right)}\left(1-\frac{2}{1+e^{-2\alpha r}}\right)$$

$$n^2 + (2n+1)\left(\sqrt{\frac{Mc^2+E_n}{\hbar^2c^2}(Mc^2-E_n-V_0)+k(k+1)(A_0+A_1+A_2)} + \sqrt{\frac{Mc^2+E_n}{\hbar^2c^2}(Mc^2-E_n+V_0+2V_3)+k(k+1)A_0}\right)$$

$$+2\left(\frac{Mc^2+E_n}{\hbar^2c^2}\right)(Mc^2-E_n-V_3)+k(k+1)(2A_0+A_1+A_2)$$

$$+2\sqrt{\left(\frac{Mc^2+E_n}{\hbar^2c^2}(-Mc^2+E_n+V_0+2V_3)-k(k+1)A_0\right)\left(\frac{Mc^2+E_n}{\hbar^2c^2}(-Mc^2+E_n+V_0)-k(k+1)(A_0+A_1+A_2)\right)} = 0 \quad (17)$$

## 3. Binding energy for Ra isotopes

In the core-cluster model, we consider Ra isotopes and solve the Dirac equation for phenomenological potential using the PNU method, and obtain the energy and wave function equation; you can see binding energy of Ra isotopes in table1.

Table1. Binding energy for Ra isotopes

| Isotope and core-cluster | $V_0$ (Mev) | $V_1$ (Mev) | $V_2$ (Mev) | $V_3$ (Mev) | $\alpha$ | R (fm)[13] | $E_{our}$ (Mev) | $E_{exp}$ (Mev)[14] |
|---|---|---|---|---|---|---|---|---|
| $^{223}Ra \rightarrow ^{209}Pb + ^{14}C$ | 2.0 | 35.0 | 2.0 | 450.0 | 0.05 | 5.6973 | 1713.29 | 1713.82 |
| $^{225}Ra \rightarrow ^{211}Pb + ^{14}C$ | 2.0 | 75.0 | 2.0 | 450.0 | 0.05 | 5.7156 | 1725.27 | 1725.20 |
| $^{227}Ra \rightarrow ^{213}Pb + ^{14}C$ | 2.0 | 125.0 | 2.0 | 450.0 | 0.05 | 5.7283 | 1740.02 | 1740.54 |

In Table 1, we assume that the Ra isotopes can be studying in the cluster model, but we know that in fact the $^{14}C$ isotope only emitted from the $^{223}Ra$ nuclei. We calculated the binding energy by choosing the best value for the potential adjustable parameters that it is very little in agreement with the experimental data.

## 4. Results and discussions

The discussion of heavy isotopes is one of the major issues in nuclear physics, and solving the N-particle problem is one of the challenges of nuclear physics and physics. In this study, we investigated the heavy isotopes of Ra, for example, $^{223}Ra$ has 223 particles, which the 223-particle problem solving virtually impossible. Therefore, scientists began modeling for multi-particle nuclei. We also used the cluster model for Ra isotopes, in the cluster model, we have to observe clusters in the experiment, in recent year's experiments in the field of heavy nuclei, the $^{14}C$ isotope emission observed.

We considered in the core-cluster model, the daughter nuclei as the core and $^{14}C$ as the cluster.

We need a potential V(r) to describe the core-cluster model, which has three essential parts, including nuclear term ($V_N$) plus Coulomb potential ($V_C$) plus centrifugal potential ($V_L$):

$$V(r) = V_N(r) + V_C(r) + V_L(r) \tag{18}$$

You can see some of these potentials worked in the past in ref [15-19]. The potential we introduce (Eq.4) is a combination of the Rosen-Morse potential [20] and a modified relativistic potential of ref [15] and an exponential type potential that incorporates the Coulomb potential itself. We take nuclei in the relativistic model so we choose the Dirac equation to obtain the nuclei wave function, so replaces the potential in the Dirac equation and obtain binding energy after solving the equation.

In this study, we calculated the binding energy for Ra isotopes with given core-cluster model respect to heavy cluster $^{14}$C and we compared our result with experimental data; the difference between them is very small.

# References

[1] H. J. Rose and G. A. Jones, Nature 307 (1984) 245.

[2]  R. Bonetti and A. Guglielmetti, in Heavy Elements and Related New Phenomena, vol. II, edited by W. Greiner and R. K. Gupta (World Scientific, Singapore, 1999) P. 643.

[3] N. Roshanbakht and M. R. Shojaei, Mod. Phys. Lett. A, Vol. 34, No. 20 (2019), 1950158.

[4] K. P. Santhosh, B. Priyanka and M. S. Unnikrishnan, Nucl. Phy. A Vol. 889, (2012), P. 29-50.

[5] D. N. Poenaru, R. A. Gherghescu, W. Greiner, Phys. Rev. C 83, (2011), 014601.

[6] N. Itagaki, A. V. Afanasjev, D. Ray, arXiv: 1906.00597 [nucl-th], (2019).

[7] B. S. Chisholm, D. Erle Nelson, H. P. Schwarcz, Science (New York, N.Y.), 216, 1131 (1982).

[8] D. N. Poenaru et al, J. Phys. G: Nucl. Phys. 10, (1984), L183.

[9] W. von Oertzen et al., AIP Conf. Proc. 1165, 19 (2009).

[10] M. Mousavi and M. R. Shojaei, Pramana – J. Phys, (2017), 88: 21

[11] M. Mousavi and M. R. Shojaei, Chin. J. Phys. (2016), 1-6.

[12] N. Roshanbakht, M. R. Shojaei, Mod. Phys. Lett. A, Vol. 34, No. 20 (2019) 1950158.

[13] I. Angeli, K. Marinova, At. Data Nucl. Data Tables99, (2013), 69.

[14] M. Wang et al, Chin Phys C, Vol.36, No.12, (2012), 1603-2014.

[15] B. Buck, A. C. Merchant, S. M. Perez, Phys. Rev. C, Vol.45, No.5, (1992),  2247-2253.

[16] Z. Binesh, M. R. Shojaei, B. Azadegan, Can. J. Phys, (2019), 1-10.

[17] D. Bai, Zh. Ren, G. Röpke, Phys. Rev. C 99,(2019), 034305.

[18] K. P. Santhosh, I. Sukumaran, Pramana – J. Phys. (2019), 92: 6.

[19] A. Soylu, F. Koyuncu, A. Coban, O. Bayrak, M. Freer, Annals of Physics, Vol.391, (2019), 263-277.

[20] N. Rosen, P. M. Morse, Phys. Rev. 42, 210 (1932).